\documentclass[12pt]{article}
\usepackage{graphicx}
\usepackage{subfigure}
\usepackage{epstopdf}

\newcommand\pubnumber{}
\newcommand\pubdate{\today}

\def\oxford{University of Oxford, Denys Wilkinson Building, Keble Road, \\ OX1 3RH, United Kingdom}

\def\Title#1{\begin{center} {\Large #1 } \end{center}}
\def\Author#1{\begin{center}{ \sc #1} \end{center}}
\def\Address#1{\begin{center}{ \it #1} \end{center}}

\newcommand\pubblock{\rightline{\begin{tabular}{l} \pubnumber\\
         \pubdate  \end{tabular}}}
\newenvironment{Abstract}{\begin{quotation}  }{\end{quotation}}

\def\beq{\begin{equation}}
\def\eeq#1{\label{#1}\end{equation}}
\def\eeqn{\end{equation}}

\def\beqa{\begin{eqnarray}}
\def\eeqa#1{\label{#1}\end{eqnarray}}
\def\eeqan{\end{eqnarray}}

\let\bar=\overbar

\def\Dslash{\not{\hbox{\kern-4pt $D$}}}
\def\dslash{\not{\hbox{\kern-2pt $\del$}}}

\def\msb{{\bar{\ssstyle M \kern -1pt S}}}

\begin{document}
\begin{titlepage}
\pubblock

\vfill
\Title{$D^0-\bar{D}{}^0$ mixing studies  with the  
  decays $D^0 \to K^0_S K^\mp \pi^\pm$}

\vfill
\Author{S. Malde and G. Wilkinson }
\Address{\oxford}
\vfill
\begin{Abstract}
We demonstrate how a time-dependent analysis of the decays $D^0 \to K^0_S K^\mp \pi^\pm$ can be used to determine the $D^0-\bar{D}{}^0$ mixing parameter $y$ with a precision that is competitive with established methods. The proposed analysis 
is an inclusive study which makes use of the measurements of the coherence factor and mean strong phase difference for these decays recently performed by CLEO.
\end{Abstract}
\vfill
\vfill
\end{titlepage}
\def\thefootnote{\fnsymbol{footnote}}
\setcounter{footnote}{0}
%


\section{Introduction}
\label{sec:intro}

In the last few years measurements have been performed at $e^+e^-$ colliders and
the Tevatron which, when taken together, reveal the presence
of mixing in the $D^0-\bar{D}{}^0$ system. These measurements have relied
on a variety of analysis strategies, including the time-dependent study
of $D^0 \to K^+\pi^-$~\cite{MIXWSKP,MIXWSKP_BABAR}, $D^0 \to K^+\pi^-\pi^0$ and 
$D^0 \to K^+ \pi^- \pi^+ \pi^-$~\cite{MIXWSKPP0} decays,
the measurement of the $D^0$ lifetime in decays to 
CP-eigenstates~\cite{MIXCPEIGEN},
and the amplitude analysis of self-conjugate final states, such 
as $K^0_S \pi^+ \pi^-$~\cite{MIXKSPP,MIXKSPP_BABAR}~\footnote{Unless stated otherwise, the charge-conjugate mode
is implicit throughout this Letter.}.
The mixing is parameterised through the dimensionless variables $x$ 
and $y$, which determine the evolution of a $D^0$ into a $\overline{D}{}^0$ meson
through off-shell and on-shell intermediate states respectively.
The current values for these parameters, averaged over all analyses,
are $x=(0.63^{+0.19}_{-0.20})\%$ and $y=(0.75 \pm 0.12)\%$~\cite{HFAG}. 
More sensitive individual measurements of 
$x$ and $y$ are needed, both to test the consistency
of the overall picture, and to search for evidence of CP-violation
in $D^0-\overline{D}{}^0$ mixing, where the existing 
constraints are very weak~\cite{HFAG}. As CP-violation in the
$D^0-\overline{D}{}^0$ system is generally considered to appear at a
negligibly small level in the Standard Model~\cite{CPVINCHARM}, any observation of
such an effect would constitute a clear signal of New Physics.

In this Letter we propose to study $D^0-\overline{D}{}^0$ mixing through
a time-dependent inclusive analysis of the decays $D^0 \to K^0_S K^- \pi^+$ and $D^0 \to K^0_S K^+ \pi^-$,
in a manner similar to that of a `wrong sign' $D^0 \to K^+\pi^-$ analysis.
Whereas the Cabibbo favoured (CF) and doubly-Cabibbo suppressed (DCS) amplitudes
in the $K\pi$  final state give rise to time integrated branching ratios that differ by a factor of $\sim 300$,
for the singly-Cabibbo suppressed (SCS) modes there is approximate equality between the two final states,
with the branching ratios of  $D^0 \to K^0_S K^- \pi^-$
and $D^0 \to K^0_S K^+ \pi^+$ being
$(3.5\pm 0.5) \times 10^{-3}$  and $(2.6\pm 0.5) \times 10^{-3}$ respectively~\cite{PDG}.  This attribute gives rise to 
several interesting and distinctive features.

Section~\ref{sec:formalism} provides details on the method, and the sensitivity of 
the analysis is assessed in Section~\ref{sec:toymc}.  Conclusions
are presented in Section~\ref{sec:conclude}.  Throughout the discussion, and by way of analogy with the CF/DCS strategy, the 
$D^0$ decay to the $K^+$ ($K^-$) final state is designated as `wrong sign'  (`right sign'), and it is assumed that the direct decay 
of the $D^0$ to the wrong sign (right sign) final state proceeds by the suppressed (favoured) amplitude.

\section{Measuring $\boldmath y$  with the decays ${\boldmath D^0 \to K^0_S K^\mp\pi^\pm}$}
\label{sec:formalism}

Adopting the notation in Ref. \cite{Blaylock}, and assuming no CP violation, the $D$ mass eigenstates, $\mid D_{1,2} \rangle$, can be written in terms of the flavour eigenstates $\mid{ D^0} \rangle$ and $\mid \bar{D}{}^0 \rangle$ as follows:
\begin{eqnarray}
\label{eq:dmass}
\mid D_{1} \rangle &=& \mid D^0 \rangle + \mid \overline{D}{}^0 \rangle, \nonumber\\
\mid D_{2} \rangle &=& \mid D^0 \rangle - \mid \overline{D}{}^0 \rangle. 
\end{eqnarray} 
The time evolution of the physical states is given by
\begin{eqnarray}
\label{eq:dtime}
\mid D_j(t)\rangle = e^{-iM_j - \frac{1}{2}\Gamma_jt} \mid D_j(t=0)\rangle,
\end{eqnarray}
where $M_j$ and $\Gamma_j$ are the masses and widths of the mass eigenstates.

We are interested in the time evolution of a sample of initially pure $\mid D^0\rangle$ mesons decaying to the wrong sign final state $\mid f \rangle$. The decay can occur directly from the $D^0$ or via mixing followed by decay from the $\bar{D}{}^0$. The total decay amplitude is the sum of the amplitude from the direct decay (the suppressed amplitude) and the amplitude from $\mid \bar{D}{}^0\rangle$ to the final state (the favoured amplitude). The final state under consideration is multi-body ($n\ge3$) and hence the amplitudes are functions of the final state kinematics, $\bf{p}$. We write the suppressed amplitude as $\mathcal{A}({\bf{p}}) \equiv \langle f_{\bf{p}}\mid H\mid D^0\rangle$ and define $\mathcal A^2= \int \mathcal{A}({\bf{p}}) \overline{ \mathcal{A}({\bf{p}})} d{\bf{p}}$. The favoured amplitude is written as $\mathcal{B}({\bf{p}}) \equiv \langle f_{\bf{p}}\mid H\mid\overline{ D}{}^0\rangle$,
with $\mathcal B^2= \int \mathcal{B}({\bf{p}}) \overline{ \mathcal{B}({\bf{p}})} d{\bf{p}}$.  
The analysis is inclusive in the sense that there is no attempt to distinguish the intermediate resonances contributing to $\mid f \rangle$.
Using the definition of the `coherence factor' $R_D$ ~\cite{COHERENCE} we can write that $\mathcal{AB}R_D e^{-i\delta_D} = \int \mathcal{A}({\bf{p}}) \overline{\mathcal{B}({\bf{p}})} d{\bf p}$, where $\delta_D$ is the average strong phase difference between the suppressed and favoured amplitudes.

Defining the following variables in the usual way:
\begin{eqnarray}
M &\equiv& \frac{1}{2}(M_1 + M_2) ,   \\
\Delta M &\equiv& M_2 - M_1 ,   \\
\Gamma &\equiv& \frac{1}{2}(\Gamma_1 + \Gamma_2), \\ 
\Delta \Gamma &\equiv& \Gamma_2 - \Gamma_1,   \\
x &=&\frac{\Delta M}{\Gamma},   \\
 y&=&\frac{\Delta \Gamma}{2\Gamma},
\end{eqnarray}
we can write the time evolution of the state $\mid D^0(t) \rangle$ which is a pure sample of $\mid D^0\rangle$ at $t=0$, using expressions~(\ref{eq:dmass}) and~(\ref{eq:dtime}) as 
\begin{eqnarray}
\mid D^0(t) \rangle = f_+(t)\mid D^0 \rangle + f_{-}(t)\mid \bar{D}{}^0 \rangle,
\end{eqnarray}
where
\begin{eqnarray}
f_+(t) &\equiv e^{-iMt -\frac{1}{2}\Gamma t}\cos \left(\frac{\Gamma t}{2}(x-iy)\right), \\
f_-(t) &\equiv e^{-iMt -\frac{1}{2}\Gamma t}i\sin \left(\frac{\Gamma t}{2}(x+iy)\right).
\end{eqnarray}
Hence the total decay amplitude to the wrong sign final state $\mid f_{\bf p} \rangle$ with kinematics $\bf{p}$ is then
\begin{equation}
\langle f_{\bf p} \mid H \mid D^0(t) \rangle = \mathcal{A}({\bf p}) f_+(t) + \mathcal{B}({\bf p}) f_-(t),
\end{equation}
and the total decay rate to the wrong sign state is then given by
\begin{eqnarray}
\Gamma[D^0 \to f] &=& \int  [\mathcal{A}({\bf p}) f_+(t) + \mathcal{B}({\bf p}) f_-(t)]\times [\overline{\mathcal{A}({\bf p})} f_+(t)^* + \overline{\mathcal{B}{({\bf p})}} f_-(t)^*] d{\bf p}, \\
 &=& e^{-\Gamma t} \bigg[\frac{\mathcal{A}^2}{4} (e^{y\Gamma t} + e^{-y\Gamma t} + 2\cos(x\Gamma t)) + \frac{\mathcal{B}^2}{4} (e^{y\Gamma t} + e^{-y\Gamma t} - 2\cos(x\Gamma t)) \nonumber \\
&+& \frac{\mathcal{AB}R_D}{4} \left( e^{-i\delta_F}\left[e^{y\Gamma t} - e^{-y\Gamma t} -2i \sin(x\Gamma t)\right] + e^{i\delta_F}\left[e^{y\Gamma t} - e^{-y\Gamma t} +2i \sin(x\Gamma t)\right]\right)\bigg]. \nonumber
\end{eqnarray}

Let us define $r_D = \mathcal{A}/\mathcal{B}$  and $y' = y\cos\delta_D - x\sin\delta_D$ and specialise to the case $D^0 \to K_S^0 K^\mp\pi^\pm$.
Under the assumption that $x\ll 1$ and $y\ll 1$, the decay rate to the wrong sign state up to orders of $x^2$ and $y^2$ becomes
\begin{eqnarray}
\label{eq:wskskpitime}
 \Gamma[D^0 \to K^0_S K^+\pi^-] & =  & e^{-\Gamma t} \left[ (r^{K^0_S K\pi}_D)^2  \,+  \,r^{K^0_S K\pi}_D R^{K^0_S K \pi}_D y_{K^0_S K \pi}'\Gamma t  \right.  \nonumber \\ 
& + & \left.
\frac{\left(1-(r^{K^0_S K\pi}_D)^2 \right)x^2 + \left(1+ (r^{K^0_S K\pi}_D)^2 \right)y^2}{4} (\Gamma t)^2  \right ]  
\label{eq:wskskpi}
\end{eqnarray}
where labels indicate the $K^0_S K^\mp \pi^\pm$ specific quantities.   
We may also consider the 
rate to the right sign state, $\mid \overline{f_{\bf p}}\rangle$. In this case, the decay amplitude  is given by
\begin{equation}
\langle \overline{f_{\bf p}} \mid H \mid D^0(t) \rangle =  \overline{\mathcal{A}({\bf p})} f_+(t) + \overline{\mathcal{B}({\bf p})} f_-(t).
\end{equation}
Once more specialising to the case $D^0 \to K_S^0 K^\mp\pi^\pm$ and assuming $x\ll 1$ and $y\ll 1$ we obtain
\begin{eqnarray}
\label{eq:rskskpitime}
 \Gamma[D^0 \to K^0_S K^-\pi^+] & =  & e^{-\Gamma t} \left[ 1 + r^{K^0_S K\pi}_D R^{K^0_S K \pi}_D y_{K^0_S K \pi}'\Gamma t  \right.  \nonumber \\ 
& + & \left.
\frac{\left(1+(r^{K^0_S K\pi}_D)^2 \right)y^2 - \left(1- (r^{K^0_S K\pi}_D)^2 \right)x^2}{4} (\Gamma t)^2  \right ].
\label{eq:rskskpi}
\end{eqnarray}

It is instructive to compare with the $D^0\to K^{\mp} \pi^{\pm}$ case.  There $r_D^{K \pi} << 1$ in contrast to the SCS decays where $r_D^{K_S^0K\pi} \sim 1$.  Furthermore, being a two body decay the coherence factor reduces to unity, and  $\delta^{K \pi}_D$ is the strong phase difference between the suppressed and favoured amplitudes.  The familiar wrong-sign $K\pi$ mixing expression is then obtained:
\begin{eqnarray}
\label{eq:kpitime}
 \Gamma[D^0 \to K^+\pi^-] & =  & e^{-\Gamma t} \left[ (r^{K\pi}_D)^2  \,+  \,r^{ K\pi}_D  y_{ K \pi}'\Gamma t  \, + \, \frac{x'^2 + y'^2}{4} (\Gamma t)^2  \right ],
\end{eqnarray}
whereas the right-sign rate has negligible deviation from an exponential decay.

Several observations can be made when assessing the potential of $D^0 \to K^0_S K^\mp\pi^\pm$ in a mixing analysis:

\begin{itemize}
\item{Recently the CLEO collaboration has presented the first measurements of the coherence factor and mean strong phase difference in  $D^0 \to K^0_S K^\pm \pi^\mp$ decays~\cite{NAIK}. Preliminary results of $R_D^{K^0_S K \pi} = 0.73 \pm 0.09$ and $\delta_D^{K^0_S K \pi} = (8.2 \pm 15.2 )^\circ$ are reported.  The relatively large value of the coherence factor means that the interference term in $y'_{K^0_SK\pi}$ in expressions~(\ref{eq:wskskpi}) and~(\ref{eq:rskskpi}) receives little dilution from the presence of the intermediate resonances. The measured value can be used as a constraint in a mixing analysis.  Furthermore, the measured value of $\delta_D^{K^0_S K \pi}$ can be used to relate $y_{K^0_S K \pi}'$ and $y$.
}
\item{
The fact that $r_D^{K^0_SK\pi} >> r_D^{K\pi}$ has several consequences when comparing $D^0 \to K^0_S K^\mp\pi^\pm$ with $D^0 \to K^\mp\pi^\pm$ as a mixing channel.  Firstly in the  $D^0 \to K^0_S K^\mp\pi^\pm$  case both the wrong sign and right sign final states have sensitivity.
On the other hand, the relative size of the term in $\Gamma t \, e^{-\Gamma t}$ is small, giving the mode lower event-by-event sensitivity to $y_{K^0_S K \pi}'$ than the $K\pi$ analysis has to $y_{K\pi}'$; furthermore the   $(\Gamma t)^2 \, e^{-\Gamma t}$ term is negligible.  Therefore $D^0 \to K^0_S K^\mp\pi^\pm$ is effectively sensitive to $y_{K^0_SK\pi}'$ alone, provided the sample is sufficiently large to fit the deviation from the purely exponential decay.
}
\item{
Experimentally, it is easier to obtain a high purity sample when selecting $D^0 \to K^0_S K^\mp\pi^\pm$  than $D^0 \to K^+\pi^-$.  This is because of the higher branching ratios, and because $D^0 \to K^+\pi^-$ receives significant contamination from $\bar{D}{}^0 \to K^+\pi^-$ decays in events where a random pion of the wrong charge  is mistaken as the `slow pion'  when reconstructing the $D^{*+} \to D^0 \pi^+$ decay chain.
}

\end{itemize}

\section{Sensitivity studies}
\label{sec:toymc}

The sensitivity of a mixing measurement using $D^0 \to K^0_S K^\mp \pi^\pm$ decays is assessed via a simple Monte Carlo study. Comparisons are made between a published wrong sign $D^0 \to K^+ \pi^-$ analysis performed at BABAR~\cite{MIXWSKP_BABAR} with 384~fb$^{-1}$ and a possible measurement using $D^0 \to K^0_SK^\mp\pi^\pm$ exploiting the same dataset.

The true decay time distributions of simulated signal data are generated according to the distributions given in expressions (\ref{eq:wskskpitime}), (\ref{eq:rskskpitime}) and (\ref{eq:kpitime}), where the values of the constants required for the simulation are given in Table~\ref{tab:const} and are taken from~\cite{HFAG}, \cite{PDG} and~\cite{NAIK}.

\begin{table}\begin{center}
\caption{Input values used in the simulation study.}\vspace*{0.1cm}
\begin{tabular}{ll}\hline \hline
\label{tab:const}
Parameter & Value \\ \hline
$\tau({D^0})$ & 0.410 ps \\
$y$ & 0.0075 \\
$x$ & 0.0063 \\
$r_D^{K\pi}$ & 0.057 \\
$\delta^{K\pi}_D$ & 22$^\circ$ \\
$r_D^{K_S^0K\pi}$ & 0.88 \\
$\delta^{K_S^0K\pi}_D$ & 8$^\circ$ \\
$R^{K_S^0 K \pi}_D$ & 0.73 \\
\hline \hline
\end{tabular}
\end{center}
\end{table}

The detector resolution is simulated by assigning each event a decay time uncertainty according to a Landau distribution with most probable value 0.16 ps and $\sigma=0.04$ ps. Values of the decay time uncertainty are restricted to those less than 0.5 ps. This description of the decay time uncertainty is similar to that described in ~\cite{MIXWSKP_BABAR}. The generated decay time is then smeared by drawing a random value from a Gaussian with a width of the event decay time uncertainty. 

Three categories of background events are considered. We assume that 75$\%$ of the background events are true $D^0$ decays where a random slow pion has been used to tag the event incorrectly.
A further 15$\%$ of the background events are assumed to be combinatoric and have a true decay time of 0. The remainder of the background is assumed to be from partially reconstructed charm decays and the time distribution is generated according to an exponential decay time with lifetime = 0.2 ps. The decay time uncertainty is assigned using the same distribution as for signal events.  The resulting total decay time distribution for background events is similar to that presented in ~\cite{MIXWSKP_BABAR}.

\begin{figure}[htbp]
\begin{center}
\subfigure[Wrong sign $K\pi$.]
{\includegraphics[width=0.46\textwidth]{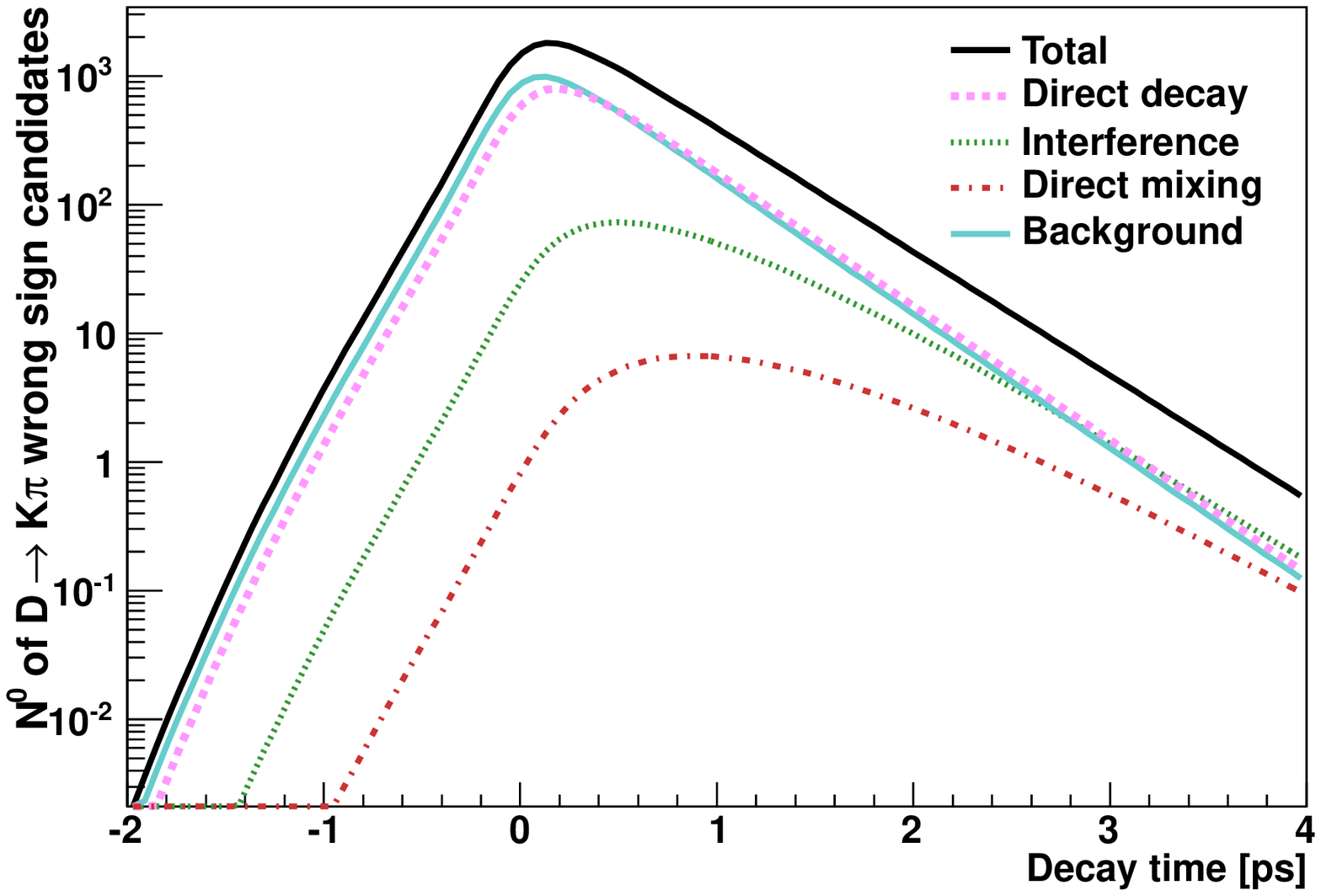}}
\qquad
\subfigure[Wrong sign $K_S^0K\pi$.]
{\includegraphics[width=0.46\textwidth]{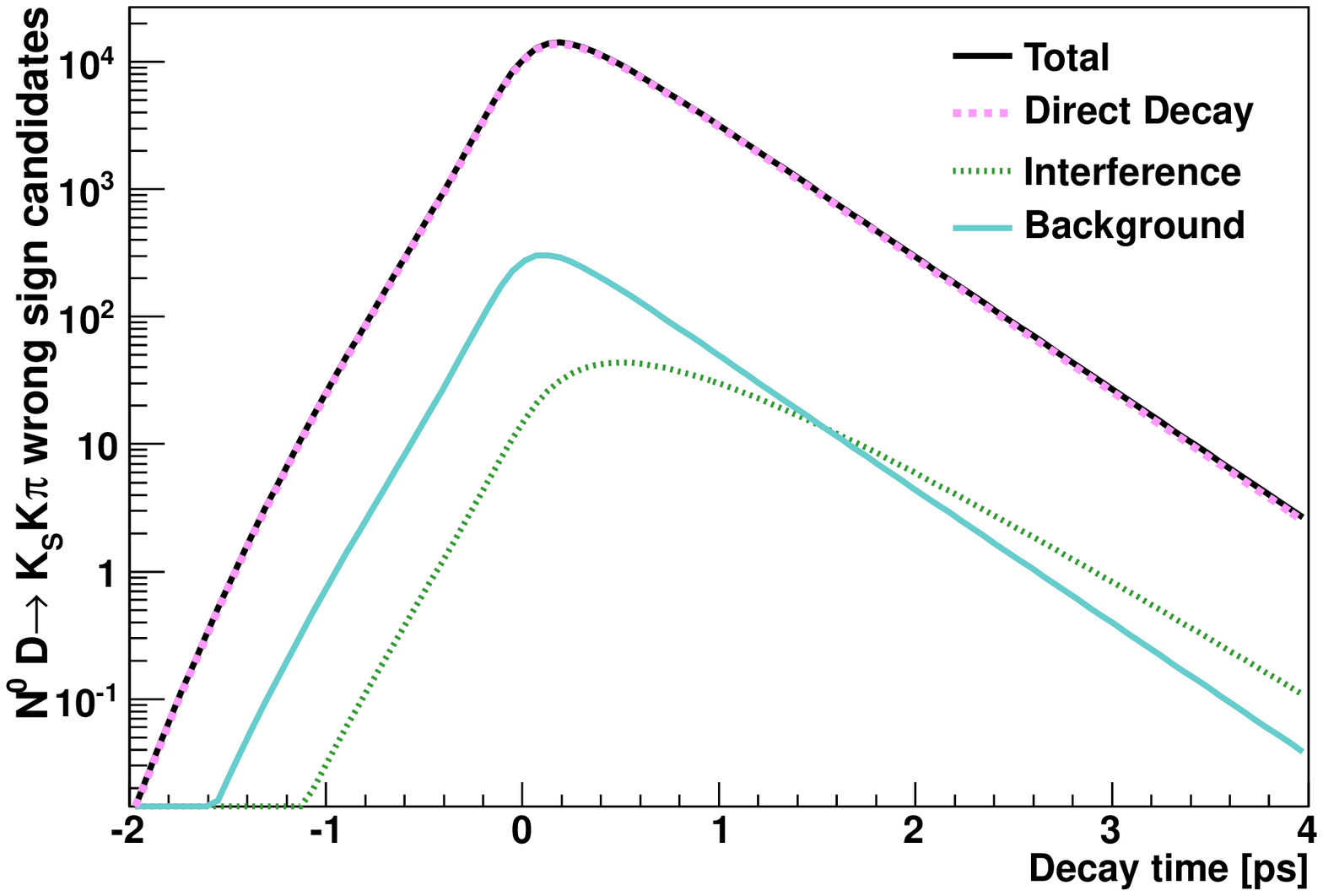}}
\caption[]{Proper time distribution of simulated candidates.}\label{fig:time}
\end{center}
\end{figure}

In 468.5 fb$^{-1}$ of data accumulated at BABAR, 540,000 tagged signal $D^0\to K_S^0 \pi^+\pi^-$ events are observed with 98.5 $\%$ purity~\cite{MIXKSPP_BABAR}. Assuming the branching fractions ratios in ~\cite{PDG} we extrapolate that the same dataset used for the $D^0\to K^{\mp}\pi^{\pm}$ mixing measurement (384 fb$^{-1}$) should yield 80,000 $D^0\to K^0_SK^{\mp}\pi^{\pm}$ signal events with purity better than 98$\%$. The signal events are split between the wrong and right sign according to their measured branching fractions.
To compare the sensitivity with existing  measurements using $D^0 \to K^+\pi^-$ decays we generate a similar Monte Carlo assuming 4000 signal events with a purity of 50$\%$ as found in ~\cite{MIXWSKP_BABAR}. The decay time distributions of the simulated wrong sign samples are shown in Figure~\ref{fig:time} where the components due to direct decay, interference, mixing, and background are shown separately. The difference in contribution of the interference term and background for the two final states is clearly visible. 

To determine the sensitivity to $y'_{K_S^0K\pi}$, 1000 independent simulated samples of events are generated and fit. A maximum likelihood fit is performed on each sample where the only floating parameter is $y'_{K_S^0K\pi}$ and all other constants are fixed to the values given in Table~\ref{tab:const} and those describing background. The likelihood ignores the second order terms in $x^2$ and $y^2$, as there are insufficient data to extract these parameters also.
The RMS spread on the returned value of  $y'_{K_S^0K\pi}$, corresponding to the expected statistical sensitivity, is 0.0054.  However the mean of the fitted values is 0.0014  lower than the input, indicating a $\sim$20\% bias.
The source of this bias has been studied. It is not a consequence of ignoring the second order terms; rather it is found to be due to an intrinsic limitation in correctly extracting the fraction of events distributed as $t e^{-t/\tau}$ in the presence of events distributed as $e^{-t/\tau}$ when the fraction is very small.  The magnitude of the bias depends on the relative size of $r_D^{K^0_S K \pi}$ and $y'_{K^0_SK\pi}$, but is found to be approximately constant within the experimentally allowed ranges of these parameters.  It is assumed that this effect can be corrected for in the measurement.

There is a  source of systematic error due to the current 12\% uncertainty in the knowledge of the coherence factor.  
This is found to induce an uncertainty of 0.0010 in the fitted value of $y'_{K_S^0K\pi}$.
Finally, it should be noted that as the time distribution of both wrong and right sign $D^0$ decays is sensitive to $y'_{K_S^0K\pi}$, the mistag background will also contain events from the interference of mixing and the suppressed decay. As the purity of the sample is high, approximating the mistag decay time by an exponential causes negligible bias.  
The total uncertainty on a $y'_{K_S^0K\pi}$ measurement using 80,000 signal decays is therefore predicted to be 0.0055 taking into account statistical uncertainties and the present knowledge of the coherence factor.

The measured value of $y'_{K_S^0K\pi}$  can be related to the mixing parameter $y$ with knowledge of $x$ and the mean strong phase difference, $\delta_D^{K_S^0K\pi}$. The sensitivity to $y$ using the known values and uncertainties of these parameters is 0.0059.

A fit study is also carried out on the generated $D^0 \to K^+\pi^-$ events. In this case the second order term in the lifetime distribution is not ignored and both $y'_{K\pi}$ and $x'^2$ are floating parameters. The correlation between the fit parameters is found to be -0.94 and the sensitivity to $y'_{K\pi}$ and $x'^2$ is 0.0046 and 0.00026 respectively. These values are very similar to those reported in~\cite{MIXWSKP_BABAR}, validating our procedure.   We can translate the result for $y'_{K\pi}$ into a measurement of $y$ and find a total uncertainty of 0.0052.
				   
\section{Conclusions}
\label{sec:conclude}

A time dependent inclusive analysis of $D^0 \to K^0_S K^\mp \pi^\pm$ decays appears a promising method to improve our knowledge of the $D^0-\bar{D}{}^0$ mixing parameter $y$.  Although the relative size of the interference effects is significantly smaller in these modes than in $D^0 \to K^+\pi^-$, compensation is provided by the higher branching ratio and expected high sample purity.  Recent preliminary results from CLEO indicate that the dilution arising from intermediate resonances to the interference term is not severe.  
The analysis can be generalised to look for evidence for CP-violating effects in $D^0 - \bar{D}{}^0$ mixing.

A sensitivity study based on existing $B$-factory publications shows that the precision obtainable on $y$ with  $D^0 \to K^0_S K^\mp \pi^\pm$ decays is similar to that of the  $D^0 \to K^+\pi^-$ analysis.  Experimentally, careful checks of the proper time determination will be required to ensure that systematic effects do not bias the measurement.  Validation could come from demonstrating a good understanding of the proper time distribution of the mixing-free decays of non-$D^0$  mesons to final states of similar topology.

\section*{Acknowledgments}

We acknowledge useful discussions with Tim Gershon and other CLEO colleagues from the United Kingdom and India.  We are grateful for support from 
the STFC, United Kingdom.

\end{document}